\documentclass[proceedings]{JHEP3}

\PrHEP{PrHEP unesp2002}                 % Do not change this one
\conference{Workshop on Integrable Theories, Solitons and Duality}

\usepackage{epsfig}                   % please use epsfig for figures
\usepackage{amsmath}
\def\rf#1{(\ref{eq:#1})}
\def\lab#1{\label{eq:#1}}
\def\nn{\nonumber \\}
\newcommand{\beano}{\begin{eqnarray*}}
\newcommand{\enano}{\end{eqnarray*}}
\def\bea{\begin{eqnarray}}
\def\ena{\end{eqnarray}}
\def\foot#1{\footnotemark\footnotetext{#1}}
\newcommand{\rg}{r_{\!g}}
\newcommand{\hg}{h_g}
\font\fld=msbm10 at 12 pt

\newcommand{\fl}[1]{\mbox{\fld #1}}     % Real, Complex, Natural etc.
\def\alv{\vec{\alpha}}
\def\bev{\vec{\beta}}

\def\lav{\vec{\lambda}}
\def\fract#1#2{\frac{#1}{#2}}

\title{Aspects of the Homogeneous Sine-Gordon models}

\author{Patrick Dorey\\
        Department of Mathematical Sciences\\
                                University of Durham\\
        Durham DH1 3LE, UK\\
           E-mail: \email{p.e.dorey@dur.ac.uk}}

\author{and \speaker{J. Luis Miramontes}\\
        Departamento de F\'\i sica de Part\'\i culas\\
        Universidad de Santiago de Compostela\\
        15782 Santiago de Compostela, Spain\\
        E-mail: \email{miramont@usc.es}}

\abstract{The physical mass scales
that determine the behaviour of general (simply-laced)
Homogeneous Sine-Gordon models are investigated
by means of a study of their finite-size
effects, using the thermodynamic Bethe ansatz. 
These models describe integrable multiparameter
perturbations of
the theory of level-$k$ $G$-parafermions, where $G$ is a Lie group.
The parameters can be related to
adjustable mass scales of stable and unstable particles.  Our results confirm
the presence of unstable particle states at generic values of
$k$, as predicted at large $k$  by semiclassical arguments.
}

\begin{document}

In this talk, we review some recent work
on the multiparameter nature of a family of
two-dimensional integrable quantum field
theories called the Homogeneous sine-Gordon
(HSG) models.
More details will be presented in~\cite{NEW},
which is currently in preparation.

\section{Introduction}
It is often productive to consider a quantum field theory as a
perturbation of its ultraviolet limit. In two dimensions this approach
is particularly effective, because the unperturbed theory is expected
to be a conformal field theory, and two-dimensional conformal
field theories are very well-understood.
The power of this approach was first fully appreciated by
A.B.~Zamolodchikov. His $c$-theorem
established non-perturbatively the
monotonic nature of renormalisation group flows in unitary
two-dimensional quantum field theories, and provided an
important tool -- the $c$-function -- for the analysis of these
flows\,\cite{CTHM}.
As a first application,
he was able to show that the perturbation of
the minimal conformal field theory ${\cal M}_{p,p+1}$
by its $\phi_{13}$ operator could induce a flow to the neighbouring
theory ${\cal M}_{p-1,p}$~\cite{INTERP}.
This result was obtained by combining the non-perturbative
$c$-theorem with the fact that the separation of the two fixed points
vanishes in the limit $p\to\infty$. The treatment, perturbative
in $1/p$\,, was in this respect
analogous to Wilson's $\varepsilon$ expansion near to four dimensions.

To go further requires a second key observation of Zamolodchikov's,
namely that some perturbations of conformal field theories break
their infinite-dimensional conformal symmetry in a way sufficiently gentle
that not all conservation laws are lost\,\cite{PCFT}. While not being scale
invariant, the resulting models should still be integrable.
With the aid of such theories one can hope to place disconnected fixed
points of the renormalisation group within a web of exactly-solvable
interpolating
theories, thereby giving a much more detailed understanding of the set
of all two-dimensional quantum field theories.

Given an integrable quantum field theory, it is natural to look for
exact, nonperturbative techniques for the analysis of its
renormalisation group flow. The exact S-matrix is insufficient for this
task, related as it is to the long-distance (on-shell)
properties of the theory. However, an S-matrix should characterise a
model completely, and once it is known exactly, there
are various ways to extract off-shell data. One of these is the
form-factor expansion of correlation functions. In principle this
allows the correlators entering Zamolodchikov's definition of his
$c$-function (or those
appearing in the integral representation derived by
Cardy\,\cite{CRDY})
to be evaluated at all
distance-scales, and once these are known the RG flow can be analysed
through the evolution of $c_{\rm zam}(r)$, as $r$ varies from $0$ to
$\infty$. However its very name gives away the fact that the
form-factor technique provides an {\em expansion} for correlation
functions, and in practice it is very hard to go beyond the first four
to six terms, even with the help of Monte Carlo
methods to evaluate the necessary multidimensional integrals.
Only very rarely (in cases which in some sense are `free' in the
relevant limit) is it possible to resum the series with sufficient
precision to characterise exactly the deep ultraviolet behaviour.

For this reason, an alternative technique, called the `thermodynamic
Bethe ansatz', or TBA, has become popular. First introduced into the
context of relativistic continuum field theory by Al.B.~Zamolodchikov
in \cite{TBAGEN}, the basic idea is to probe a theory via its
so-called effective central charge $c_{\rm eff}(r)$, which is defined
as follows. Let $M$ be a mass scale in the perturbed theory -- either
the mass of an asymptotic state, or the inverse of a crossover scale
if the theory is again massless in the far infrared.
Now suppose that the one-dimensional space on which the model
is defined is rolled up into a circle. The
spectrum becomes discrete and all energy levels depend on the
circumference $R$ of the circle. In particular, this is true
for the ground state energy, which on general grounds should have
the form
\begin{equation}
E(R)={\cal E}R-\frac{\pi}{6R}\,c_{\rm eff}(M\!R)
\label{ceffdef}
\end{equation}
where ${\cal E}R$ is a `bulk' contribution proportional to the system
size, and the more subtle parts of the $R$-dependence have
been wrapped up in the dimensionless function $c_{\rm eff}(M\!R)$, the
`effective central charge'. The name comes from the fact that, if a
theory {\it is} conformal, then $c_{\rm eff}$ is a constant, equal (in
unitary theories) 
to the central charge of the theory\,\cite{CEFF}.
Thus $c_{\rm eff}(r)$ has most of the useful properties of Zamolodchikov's
$c$-function $c_{\rm zam}(r)$\,, though it is important to appreciate that
away from the conformal points, the two functions are not the same.
Since we'll be concentrating on the effective central charge in this
article, from now on we will simply denote it as $c(r)$.

A definition such as (\ref{ceffdef}) is all very well, but without a
practical way to calculate $c(r)$, it is
of little practical use. Fortunately Al.B.~Zamolodchikov has shown how
to apply  general ideas of the thermodynamics of systems solvable by the
Bethe ansatz to relativistic systems for which an exact S-matrix is
known\,\cite{TBAGEN}. The upshot is a set of nonlinear integral
equations, now called  `TBA equations'. Once these equations have been
solved for a set of functions known as pseudoenergies, the effective
central charge can be extracted as an integral.  Specific
examples will be seen later in the paper.  While
exact solutions of the TBA equations
are hard to come by, their numerical treatment 
is very straightforward, and limiting behaviours can be
obtained analytically.
To give the reader a flavour  of the results that can be
obtained using the TBA method, in the remainder of this section
we sketch some of the simple and not-so-simple behaviours that have
been found over the last few years.

First of all, if a model is massive in the far infrared, then
the $1/R$ `correction' term to the bulk behaviour of
(\ref{ceffdef}) must be absent in that limit, and so $c(r)$ must tend
to zero as $r\to\infty$. Conversely, in the deep
ultraviolet $c(r)$ will tend to the central charge $c_{UV}$ of the
unperturbed conformal field theory, and in the most simplest
situations $c(r)$ just interpolates between these two values as
$\log(r)$ varies from $-\infty$ to $+\infty$, as illustrated in
figure~\ref{flow1}. This was the behaviour seen in
\cite{TBAGEN} for the Lee-Yang and three-state Potts models;
subsequently, the method has been applied to many other theories.

\FIGURE[h]{
\epsfig{file=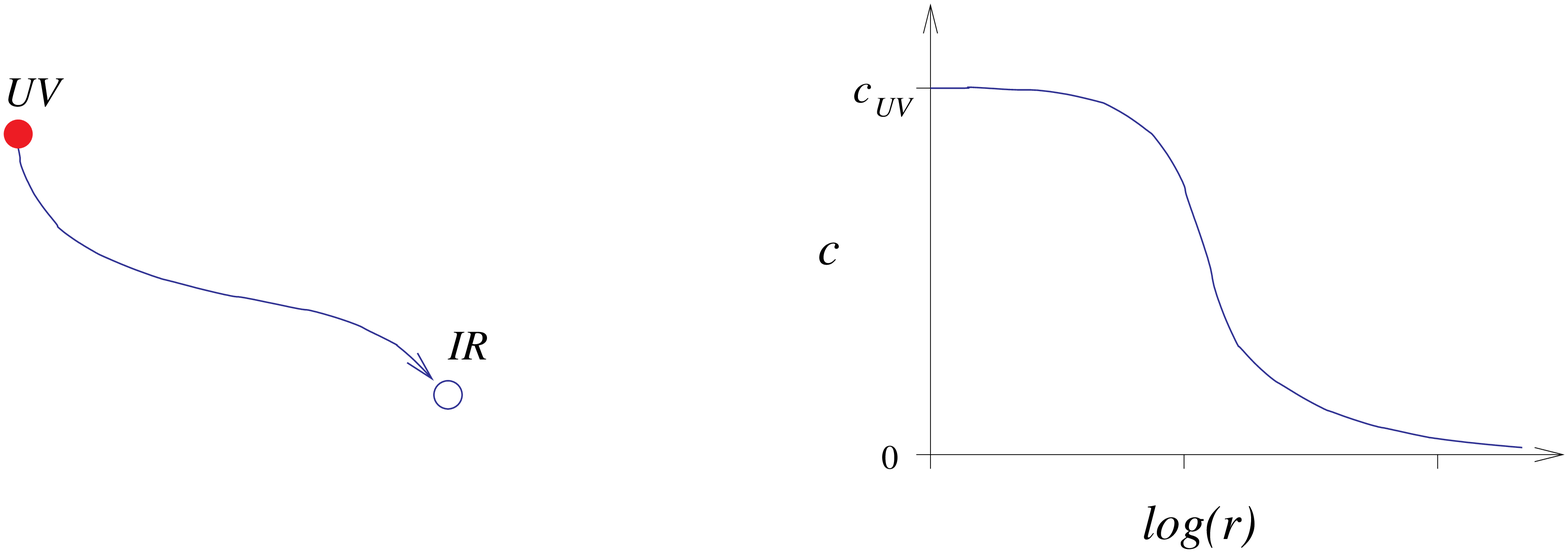,width=0.8\linewidth}
\caption{The RG flow and effective central charge for a typical
massive perturbed conformal field theory. The filled dot denotes a
fixed point of the RG with infinite correlation length; the empty dot,
a `massive' fixed point with zero correlation length.}
\label{flow1}
}

For models with massless infrared limits, which interpolate between
conformal field theories in the manner of the $\phi_{13}$
perturbation of ${\cal M}_{p,p+1}$
already mentioned above, the TBA machine is a little harder to use.
The first examples of massless TBA systems were conjectured, for
the interpolating $\phi_{13}$ flows, by
Al.B.~Zamolodchikov in \cite{MSSLSS}; they served to give
nonperturbative confirmation that the picture of the flows developed
by A.B.~Zamolodchikov
for large $p$ holds in general.
Again, the TBA equations yield a function $c(r)$ interpolating between
the expected values, in these cases the central charges
$c_{UV}$ and $c_{IR}$ of the ultraviolet and infrared limiting models.
The general picture is shown in figure \ref{flow2}.

\FIGURE[h]{
\epsfig{file=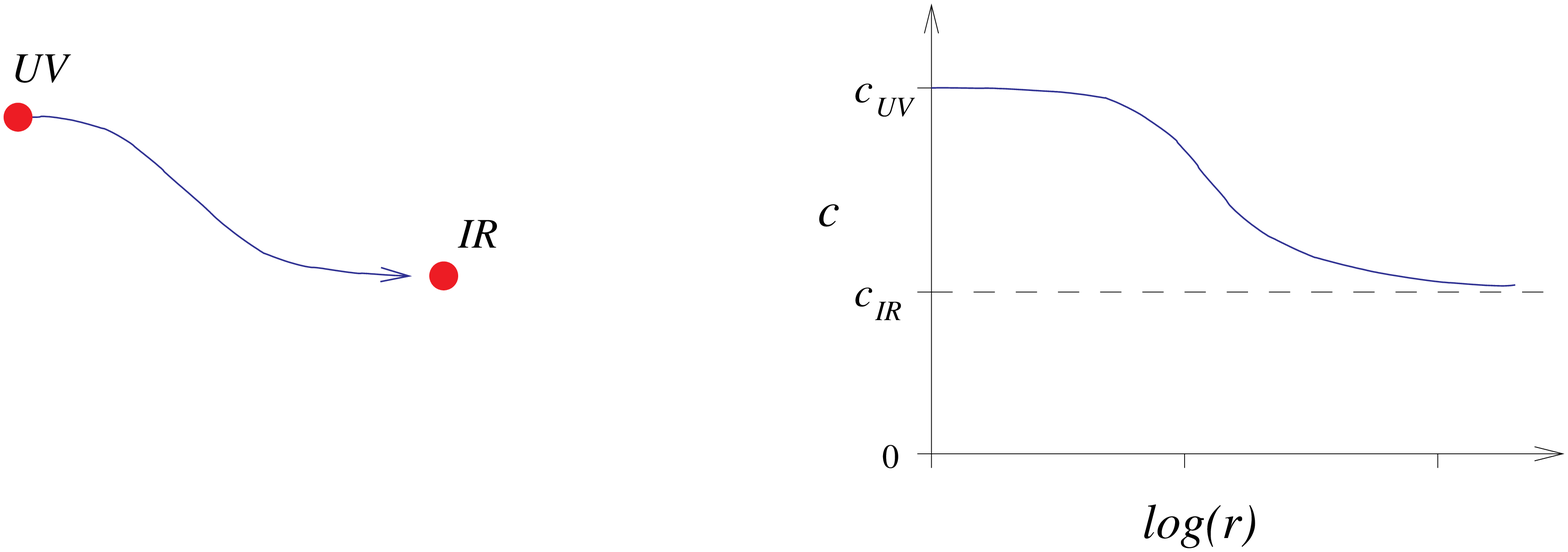,width=0.8\linewidth}
\caption{
A typical massless flow.
Labelling conventions as in figure \ref{flow1}.
}
\label{flow2}
}

The next development was less expected. In \cite{ZAMO}, Al.B.\
Zamolodchikov pointed out a particularly simple TBA system, related by
analytic continuation to that of the sinh-Gordon model, for which the
flow pattern of the effective central charge was as illustrated on the
right-hand side of figure \ref{flow3}.

\FIGURE[h]{
\epsfig{file=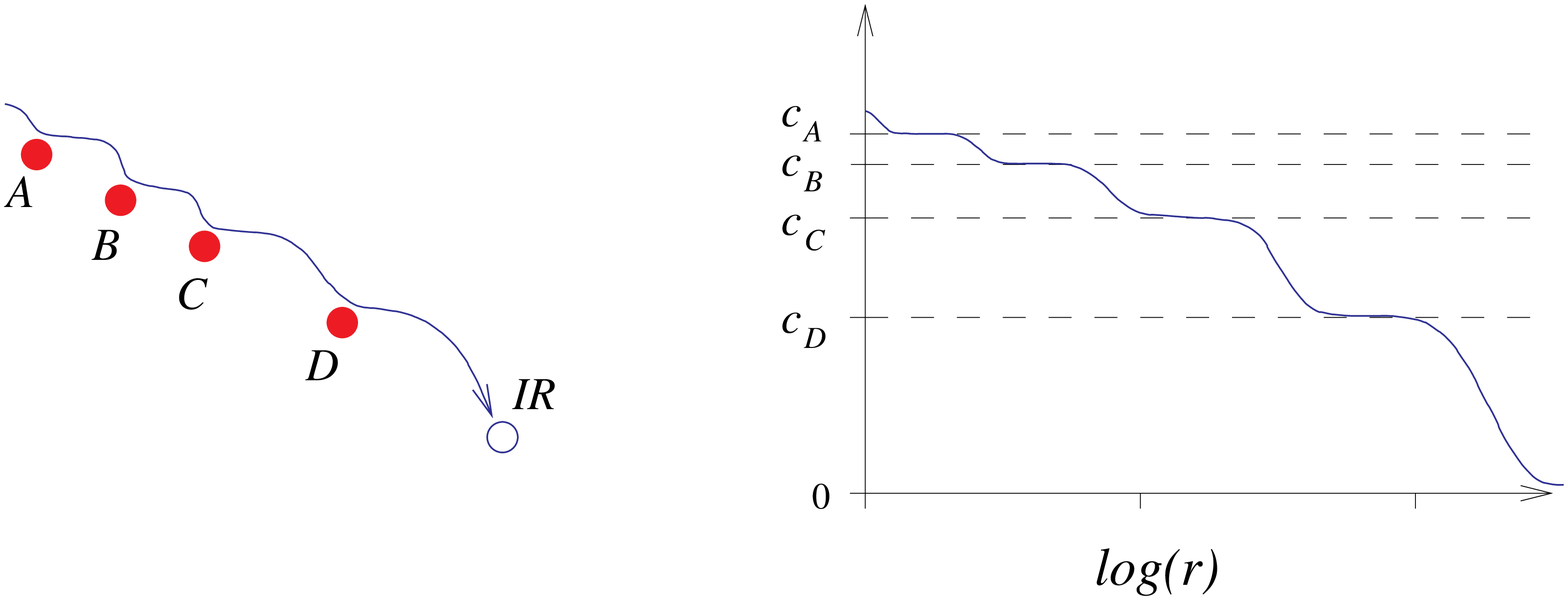,width=0.8\linewidth}
\caption{
A (massive) `staircase' flow,
passing close to the RG fixed points $A$, $B$, $C$
and $D$.
Labelling conventions as in figure \ref{flow1}.
}
\label{flow3}
}

Over an infinite set of intermediate distance ranges,
the effective central charge stabilises at an approximately constant
`plateau' value. For the
system introduced in \cite{ZAMO}, the plateau values are precisely
the central charges of the minimal models ${\cal M}_{p,p+1}$\,; the
natural interpretation is that the effective central charge produced
by Zamolodchikov's  TBA is that of an integrable quantum field theory
whose RG trajectory passes close by the fixed points of all of the
minimal models in turn, as shown on the left-hand half of
figure~\ref{flow3}.  The conjectured theory is called a staircase
model, for reasons which should be obvious on examining the figure.
Subsequent work has generalised this in a number of directions
\cite{PAT,MARTINS,SPIRAL}, showing in particular
\cite{SPIRAL} that for some staircase models, the IR limit is massless.
A couple of other properties are worth noting: first, the treatment of
the deep UV limits of staircase models is rather subtle; and second,
each staircase model in fact sits in a one-parameter family of such
models, the parameter controlling how closely each fixed point is
visited on the journey from ultraviolet to infrared. This extends the
idea, alluded to above,
that there should be a `web' of integrable models linking the RG
fixed points -- in many cases,
the trajectories making up this web are now seen to be
just the boundaries of two-dimensional surfaces of integrability in
the space of theories, swept out by the one-parameter families of
staircase models.

Finally, we come to the sets of flows which will be the focus of our
attention in the rest of this article. The homogeneous sine-Gordon (HSG)
models will be introduced in greater detail below. It turns out that
their RG behaviours combine various features of the flows discussed above
in an interesting way. On the one hand, their deep UV limits are
straightforwardly described as coset conformal field theories, with
none of the subtleties seen in the cases of the staircase models. On
the other hand, for suitable choices of the perturbing parameters,
they can pass close by a number of other conformal field theories on
their way to the infrared. Figure \ref{flow4} shows one possible
scenario -- note that, by a suitable tuning of the parameters, it has
been arranged that the fixed point $B$ has been missed, so only the
conformal fixed points $A$, $C$ and $D$ are seen by the flow.

\FIGURE[h]{
\epsfig{file=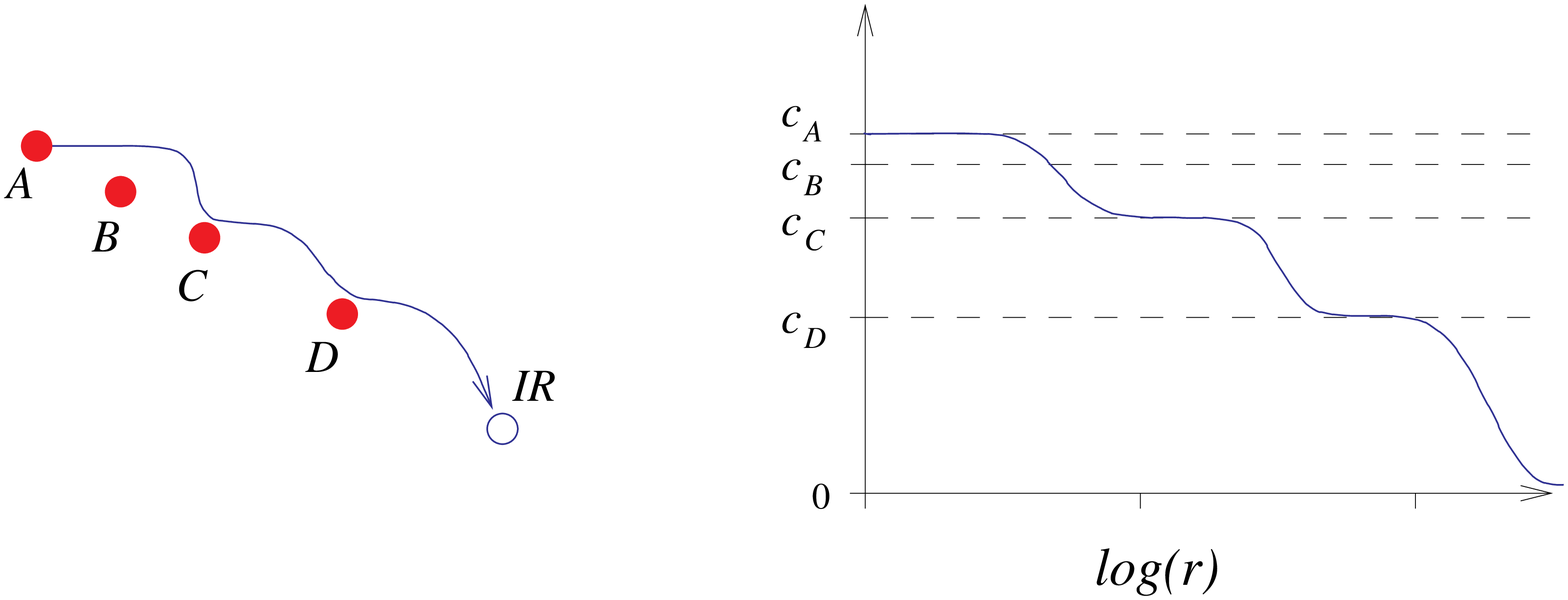,width=0.85\linewidth}
\caption{
An HSG flow. 
The ultraviolet fixed point is $A$\,, so $c_{UV}=c_A$\,.
Since the flow remains far from the fixed point $B$, there is no plateau
in $c(r)$ at $c_B$.
Labelling conventions as in figure \ref{flow1}.
}
\label{flow4}
}

Since the HSG models have many parameters, there are many options for
the flows, and considered together they sweep out whole manifolds of
integrability in the space of theories.
Our main aim in \cite{NEW} was to understand the
possibilities in detail, and to relate them to the known properties of
the models in their semiclassical limits. The TBA treatment of
HSG models had previously been discussed in \cite{HSGTBA}, while an
analysis of the flows via Zamolodchikov's $c$-function was given
in \cite{FFRG,FF2}. However, we believe that the RG behaviour of
general HSG models is considerably richer than had been suspected on
the basis of the earlier work. In the rest of this article we shall
outline some aspects of this story.

\section{The Homogeneous Sine-Gordon models}

After the rather discursive introduction, we now get down to
specifics.  The HSG models~\cite{HSG}
are two-dimensional integrable quantum field theories
that describe integrable perturbations of coset
conformal field theories (CFTs) of the form $G_k/U(1)^{\rg}$, where $G$ is
a simple compact Lie group with Lie algebra $g$, $k>1$ is an integer,
and $\rg$ is the rank of $g$. Equivalently, they are
integrable perturbations of the theory of level-$k$
$G$-parafermions\,\cite{GEPNER}, and can be viewed as the generalization
of the perturbation of the ${\fl Z}_k$ parafermionic CFT by its first
thermal operator\,\cite{PARAINT}. The latter is recovered for
$G=SU(2)$~\cite{BAKAS}, and is described by the minimal $a_{k-1}$
factorised $S$-matrix\,\cite{MINS}.

For $G\not=SU(2)$, the HSG models are examples of multiparameter
deformations of CFTs. From this perspective, the models can
be specified by actions of the form
\begin{equation}
S_{\rm HSG}=
S_{\rm CFT}-\mu\int d^2x\, \Phi\,,
\lab{PCFTaction}
\end{equation}
where $S_{\rm CFT}$ denotes an action for the CFT of
level-$k$ $G$-parafermions, and
$\Phi$
is a spinless primary field with conformal dimensions
$\Delta_{\Phi}= \bar\Delta_{\Phi}= \hg^{\vee}/(k+\hg^{\vee})$, with
$\hg^{\vee}$ the dual Coxeter number of $g$. The field $\Phi$ lies in a
multiplet in the unperturbed CFT, and a particular field $\Phi$ is
determined by $2\rg-2$ dimensionless parameters. Thus, in addition to the
dimensionful parameter $\mu$, means that the theory depends on
$2\rg-1$ parameters, one of which can always be mapped onto the overall
scale.

A more explicit definition of the models corresponding to
perturbations of the coset $G_k/U(1)^{\rg}$ (which we shall refer to
as the $G_k$--HSG in the
following) is provided by the action~\cite{HSG}
\begin{equation} 
S_{\rm HSG}[{\gamma}]= S_{\rm gWZW}[{\gamma}] \>-\int d^2x\>
V({\gamma})\>,
\lab{Action}
\end{equation}
where ${\gamma}={\gamma}(t,x)$ is a bosonic field taking values
in the group $G$, $S_{\rm gWZW}$ is the gauged
Wess-Zumino-Novikov-Witten (WZW) action corresponding to the coset
$G/U(1)^{\rg}$~\cite{GWZW}, and the potential is
\begin{equation} 
V({\gamma}) =\frac{m_0^2}{4\pi\beta^2} \> \langle
\Lambda_+ , {\gamma}^\dagger \Lambda_- {\gamma}\rangle\>.
\lab{Potential}
\end{equation}
Here, $\langle\;,\>\rangle$ is the Killing form of $g$, $m_0^2$ and
$\beta^2$ are the bare mass scale and coupling constant, respectively, and
$\Lambda_\pm$ are two arbitrary
elements in the Cartan subalgebra of $g$ associated with the maximal torus
$U(1)^{\rg}$, chosen to be not orthogonal to any root of $g$. Notice
that the freedom to choose $\Lambda_\pm$ involves precisely
$2\rg-1$ adjustable parameters. In the quantum theory, the coupling
constant becomes quantised,
$\beta^2\simeq 1/k + O(1/k^2)$, and its role is played by the integer
number $k>1$ known as the `level'.

In this context, the $k\rightarrow\infty$ limit corresponds to both the
weak-coupling (perturbative) and semiclassical regimes of~\rf{Action},
and this allows the $G_k$--HSG models to be analysed at large $k$ using
semiclassical techniques~\cite{SEMI}. This has lead to conjectures for
the mass spectra and S-matrices at arbitrary values of
$k$ for $G$ simply-laced~\cite{SMAT} and non simply-laced~\cite{KORFF}. 
In the simply-laced cases, the S-matrices have now
been checked using both TBA~\cite{HSGTBA} and
form-factor~\cite{FFRG,FF2,FF1} approaches, and there is little doubt that
they are correctly describing the perturbed parafermionic theories.
However, there are aspects of these models that still have to be
clarified.

One of them is the presence of unstable particles, which is a conjecture
that also emerges from the semiclassical studies. Indeed, for the 
simply-laced $G_k$--HSG model, a total of $k-1$ particle-like
states were identified for each positive root of $g$
in~\cite{SEMI}, but only those corresponding to the roots of height~$1$
-- the simple roots -- turn out to be stable. In S-matrix
theory, stable and unstable particles play completely different
roles~\cite{EDEN}, and it is only the particles associated with the simple
roots whose existence is immediately confirmed by the checks on the exact
S-matrices. Fortunately, there are physical observables, such as
correlation functions and finite-size effects, where both types of
particles are expected to play similar roles, setting the scales of
crossover phenomena. This is because the effective
behaviour of the system at a certain energy scale depends on
the number of particle states which are effectively massless compared to
that scale, irrespective of their stability.
Examining the system at different
scales thus 
provides a well-defined method to detect the existence of
physical mass scales associated with
both stable and unstable particles: at the mass scales corresponding to
a physical particle state, a sharp change in the behaviour of the
system should be observed, due to the decoupling of that
particle.

The study of crossover phenomena for certain HSG models has already
provided non-perturbative evidence of the existence of physical mass scales
associated with the roots of heights~1 and~2. In
particular, the TBA was applied to the study
of finite-size effects in the $SU(3)_k$--HSG models in~\cite{HSGTBA}.
Moreover,
the renormalisation group flow of Zamolodchikov's
$c$-function was calculated in~\cite{FFRG,FF2} for the
$SU(N)_2$--HSG models, making use of the expansion of
two-point correlation functions in terms of form factors. However,
in~\cite{FF2}, no trace was found of any mass scale associated with the
roots of height larger than~2.

The main result of~\cite{NEW} is that the study of
finite-size effects using the TBA technique indeed provides
non-perturbative  evidence for the presence of unstable particles states
associated will {\em all} the roots of $g$, and at any value of $k$, large
or
small. Nevertheless, due to the nature of crossover phenomena, the study
of finite-size effects does not allow to find the value of the
corresponding mass scales with arbitrary precision; the most that one can
ask is to pick up {\em well-separated} scales. Therefore, in order to
understand properly the results of the TBA analysis, it is
convenient to have a clue as to how many separated scales can be expected.
In our case, this is provided by studying how many separated classical
mass scales can be manufactured by varying the $2\rg{-}1$
adjustable parameters; this will be addressed in
the following section.

\section{Particle masses and classical mass scales}

The $S$-matrices proposed in~\cite{SMAT} for the simply-laced HSG models
describe the scattering between solitonic particles labelled by two quantum
numbers, $(i,a)$, where $i=1\ldots\rg$ and
$a=1{\ldots}k-1$ for the $G_k$-HSG models. The
mass of the particle $(i,a)$ is
\begin{equation}
M_a^i = M m_i\> {\mu}_a\>,
\lab{Perron}
\end{equation}
where $M$ sets the overall mass scale,
$m_i$, $i=1\ldots \rg$, are $\rg$ arbitrary (non vanishing)
relative mass scales attached to the nodes of the Dynkin diagram of $g$,
and $\mu_a=\sin (\pi a/k)$ are the components of the Perron-Frobenius
eigenvector of the $A_{k-1}$ Cartan matrix. In addition to the mass
ratios of the stable particles,  the S-matrix is sensitive to a
further $\rg-1$ so-called resonance
parameters $\sigma_{ij}=-\sigma_{ji}$, initially
defined for each pair
$\langle i,j\rangle$ of neighbouring
nodes on the Dynkin diagram of $g$. These are most conveniently
specified by assigning a variable $\sigma_i$ to each node
of $g$ and setting $\sigma_{ij}=\sigma_i-\sigma_j$. This way, the
$G_k$-HSG factorised $S$-matrix theory is determined by the $2r_g-1$
parameters $\{m_i,\sigma_{ij}\}$.

Via the index $i$, each set of $k{-}1$ stable particles in
the quantum theory is associated with a simple root $\vec\alpha_i$.
In addition, classically, the theory exhibits sets of solitonic
particle-like solutions associated with all of the other positive roots
$\vec\beta\in\Phi^+$ as well. Their
mass scales can be written in a concise way as follows. Ignore for a
moment the fact that $m_i$ and
$\sigma_{ij}$ are parameters of the quantum theory, and define a
couple of vectors in the weight  space of $g$ by setting
\begin{equation}
\vec\lambda_{\pm}=\sum_1^{\rg}m_i\> e^{\pm\sigma_i}\>
\vec\lambda_i
\lab{LambdaW}
\end{equation}
where the $\vec\lambda_i$\,, $i=1\dots \rg$\,, are the fundamental
weights of $g$ and satisfy $\vec\lambda_i\cdot \vec\alpha_j=\delta_{ij}$.
Then the relative mass scale of the particles
associated with the positive root $\vec\beta$ is given by the formula
\begin{equation}
m^2_{\vec\beta}=(\vec\lambda_+\cdot
\vec\beta)\,(\vec\lambda_-\cdot\vec\beta)~,
\lab{MassRoot}
\end{equation}
which reduces to $m_i^2$ for $\vec\beta=\vec\alpha_i$, as it should.

Eqs.~\rf{LambdaW} and~\rf{MassRoot} allow one to determine the number of
well-separated (classical) scales that  can be constructed by tuning the
adjustable parameters. Take a positive root $\bev=\sum_1^{\rg} c_i
\alv_i$, and substitute this expansion in~\rf{MassRoot}. Then,
\begin{equation}
m^2_{\vec\beta}=
\sum_{i,j=1}^{\rg}c_ic_j\> m_im_je^{\sigma_i-\sigma_j}
\lab{MassRootb}
\end{equation}
shows that all squared masses are expressed as linear combinations
of the $\rg(\rg{+}1)/2$ quantities
\begin{equation}
2m_im_j\cosh(\sigma_i{-}\sigma_j),\quad i,j=1\dots
\rg\,,
\lab{MassRootc}
\end{equation}
with coefficients $c_ic_j$ that are independent of the parameters
$\{m_i,\sigma_j\}$, and are the squares of numbers of order one.
Therefore, the model certainly has no more than $\rg(\rg{+}1)/2$
separable mass scales, given by the numbers
\begin{equation}
m_{ij}=\sqrt{m_im_j}\,e^{|\sigma_i-\sigma_j|/2}~,\quad i,j=1\dots
\rg\,.
\lab{Mdef}
\end{equation}
Notice that only in the $SU(N)_k$--HSG theories, for which
$N=\hg=\rg{+}1$, is
$\rg(\rg{+}1)/2$ equal to the number of positive roots -- in all
other cases it is smaller. The number of separable mass scales that a
classical HSG model can exhibit is therefore generally less than the
number of positive roots. However, the scales~\rf{MassRootc} only ever
appear through the linear combinations~\rf{MassRootb}, and it could be
that some numbers from the set provided by~\rf{Mdef} never occur as the
largest term in these sums for any choice of the values of the parameters,
but rather are always swamped by other terms. Indeed, as explained
in~\cite{NEW}, this is always the case for $g\not=a_n$. For all simply-laced
Lie algebras, the maximal number of separable classical mass scales is
given in table~\ref{MaxNumb}.

\TABULAR[ht]{|c||l|}{
\hline 
\quad $g$~\qquad & \qquad Maximal number of
separable scales~\qquad~\qquad
\\
\hline \hline
\quad $a_n$~\qquad & \quad $n(n+1)/2$ \\
\hline
\quad $d_n$~\qquad & \quad $n(n+1)/2-1$ \\
\hline
\quad $e_6$~\qquad & \quad $n(n+1)/2-2\>, \quad n=6$ \\
\hline
\quad $e_7$~\qquad & \quad $n(n+1)/2-2\>, \quad n=7$ \\
\hline
\quad $e_8$~\qquad & \quad $n(n+1)/2-3\>, \quad n=8$ \\
\hline}
{\label{MaxNumb}
The maximal number of separable scales for the HSG models
associated with the different simply-laced Lie groups.}

\section{Detecting mass scales with the TBA equations}

We may now return to the quantum theory and analyse the scales
generated  in the TBA equations. It turns out that they show a very neat
match with the separable classical scales. This leads to the conclusion
that, in the quantum theory, there is also a mass scale associated with
each positive root of $g$ and, remarkably, that it is given by
Eqs.~\rf{LambdaW} and~\rf{MassRoot} as a function, now, of the {\em
quantum} $S$-matrix parameters.

Let us sketch the main features of the TBA analysis
presented in~\cite{NEW}. The TBA equations for the
HSG model have the standard form for a diagonal scattering theory,
though care is needed in their derivation owing to the parity-breaking
of the model~\cite{HSGTBA}.
There is a pseudoenergy $\epsilon_a^{i}(\theta)$ for each of the
$(k{-}1)\times \rg$ stable particles, while
the $\rg$ mass scales $m_i$ influence the equations via
$(k{-}1)\times \rg$ energy terms
\begin{equation}
\nu_{a}^i(\theta)=
 M_{a}^i\> R\> \cosh\theta =
 m_{i}\mu_a\> r\> \cosh\theta
\lab{EnergyTerms}
\end{equation}
where
$\mu_a=\sin(\pi a/k)$ as before, and we have introduced a
dimensionless overall crossover scale by
\begin{equation}
r=MR\,.
\end{equation}
Defining $L_{a}^i (\theta)=
\log (1+{\rm e}^{-\epsilon_{a}^i (\theta)})$, the system of
TBA equations to be satisfied by the pseudoenergies is
\begin{equation}
\epsilon_{a}^i (\theta)
= \nu_{a}^i(\theta) -
\sum_{b=1}^{k-1}\left(\phi_{ab}\ast L_{b}^i (\theta) +
\sum_{j=1}^{\rg}
\>I^g_{ij}\>
\psi_{ab}\ast L_{b}^j (\theta-\sigma_{ji})
\right)\>.
\lab{TBAGen}
\end{equation}
where the symbol~`$\ast$' denotes
the usual rapidity convolution, $I^g$ is the incidence matrix of $g$, and
the definition of the TBA kernels can be found in~\cite{HSGTBA}. They are
of the same form as the kernels defining other TBA systems which had
previously arisen in the contexts of perturbed coset theories
\cite{MSSLSS,RAVA}, restricted solid-on-solid models~\cite{KN}, and staircase
models
\cite{PAT,SPIRAL,MARTINS}. The effective central charge
$c(r)$ is then expressed in the standard way in terms of the energy terms
and the solutions to the TBA equations:
\begin{equation}
c(r)= \frac{3}{\pi^2}\> \sum_{i=1}^{\rg}\> \sum_{a=1}^{k-1}\>
\int_{-\infty}^{+\infty} d\theta\> \nu_{a}^i(\theta) \> L_a^i(\theta)\>.
\end{equation}
Note that, in this multiparameter case, $c(r)$ depends not only on $r$,
but also on the relative mass scales and the resonance parameters:
$c(r)=c(r; m_i,\ldots,m_{\rg},\sigma_1,\ldots,\sigma_{\rg})$.

The limiting value of $c(r)$ in the
$r\rightarrow0$ limit corresponds, at least in unitary cases such as
these, to the central charge of the
conformal field theory describing
the deep UV limit of the theory.
For the HSG theories this quantity was calculated
in~\cite{HSGTBA}, with the result
\begin{equation}
\lim_{r\rightarrow0} c(r)= \frac{k-1}{k+\hg}\> \hg \rg\>,
\lab{ConstantTBA}
\end{equation}
where $\hg$ is the Coxeter number
of $g$\,. This holds for any fixed choice of the relative mass
scales $0<m_i<+\infty$ and the
resonance 
parameters $-\infty<\sigma_{i}<+\infty$. In other words, $c(r)$
tends to the central charge of the $G_k/U(1)^{\rg}$ coset CFT in
the deep UV limit. In the opposite, $r\to\infty$, limit, $c(r)$ tends to
zero, as expected for a massive theory.

However, this is not the only information hidden inside
the TBA equations. For intermediate values of $r$, depending on the
values taken by the parameters $\{m_i\}$ and $\{\sigma_{i}\}$,
the scaling function $c(r)$ can show
the characteristic staircase pattern mentioned in the introduction,
indicating a renormalisation group
flow which passes close to a number of other fixed points.
In contrast to Zamolodchikov's original staircase model~\cite{ZAMO} and
its generalizations~\cite{PAT,MARTINS,SPIRAL}, the number of steps turns
out to be finite. Furthermore, for the HSG models all aspects of the
staircase pattern can be understood physically, as a consequence of the
decoupling of those particles that are effectively heavy at the relative
energy scale fixed by
the temperature $R^{-1}$, be they stable or unstable.
This was already observed for the $SU(3)_k$--HSG models in
\cite{HSGTBA} but, since $a_2$ only has roots of height~1 and~2,
these cases were too simple for our purposes.
The analysis presented in~\cite{NEW} applies to generic (simply-laced) 
$G_k$-HSG models, and follows a line of argument used originally
to study the staircase models in \cite{PAT,SPIRAL}. Although, for finite
$r$, the TBA equations cannot be solved  exactly, this method allows a
full understanding of the staircase pattern to be gained, subject only to
some mild assumptions about the form of the solutions to~\rf{TBAGen}.
In fact, these assumptions are no more severe than those usually made in
the analysis of the UV limit of standard TBA systems. In any case,
the resulting predictions have been checked numerically in a number of
particular cases.

The picture derived in~\cite{NEW} is based on the following
properties of the TBA equations. First of all, due to the dominance of the
energy terms
$\nu_a^i(\theta)$,
\begin{equation}
L_a^i(\theta) \approx0\
\quad {\rm for}\quad |\theta|\gg -\log(\fract{2}{m_ir})\>, \quad
a=1\dots k-1\,.
\lab{Hypo}
\end{equation}
Second, the kernels $\phi_{ab}(\theta)$ and $\psi_{ab}(\theta)$ are in a
well-defined sense {\em local}~\cite{PAT}; {\it i.e.\/}, for real values
of $\theta$, they are peaked about $\theta=0$, and fall off exponentially
to become effectively zero outside a region of order one.
This implies that the pseudoenergy $\epsilon^i_a(\theta_0)$ is influenced
by the value of the energy term $\nu_a^i(\theta)$ at
$\theta=\theta_0$, and by the pseudoenergies $\epsilon^j_b(\theta)$
at $\theta\approx\theta_0-\sigma_{ji}$\,,
for all $j$ such that $I^g_{ij}\neq 0$.

An immediate consequence of eq.~\rf{Hypo} is that the pseudoenergies
$\epsilon_a^i(\theta)$ do not contribute to the value of
$c(r)$ for
$2r^{-1}\ll m_i$.  Actually, in this regime, the TBA system corresponding
to the $G_k$-HSG models truncates to the system associated to the
$G^{[i]}_k$-HSG model, where we denote by $G^{[i]}$ the (semisimple)
Lie group associated with the Dynkin diagram obtained by removing the node
$i$ from the Dynkin diagram of $g$. Therefore, and provided that the
mass scale $m_i$ is well-separated from any other scale in the system,
the effective behaviour changes at $2r^{-1}\approx m_i$. Physically, it
corresponds to the decoupling of all the particles (stable and unstable)
associated with the roots
$\bev$ such that $\bev\cdot\lav_i\not=0$.

The remaining expected changes in the value of the finite finite-size
scaling function are found by studying
whether the shifted convolution terms in the TBA equations~\rf{TBAGen}
can be neglected of not for particular values of~$r$, again as a
consequence of eq.~\rf{Hypo}. The result can be summarised as
follows~\cite{NEW}. Take two arbitrary nodes $k\not=l$ on the Dynkin
diagram of $g$, and consider the unique chain of nodes joining them; {\it
i.e.\/},
\begin{equation}
\{i_p\>,\; p=1\ldots n\}\>, \quad {\rm with}\quad i_1=k,\; i_n=l\>,
\end{equation}
and $\{i_p,i_{p+1}\}$ being neighbouring nodes for
$p=1\ldots n-1$. Then, the
values of the pseudoenergies $\epsilon^k_a(\theta)$ and
$\epsilon^l_b(\theta)$ are independent, for any $a,b=1\ldots k-1$, when
the value of $r$ is in the range
\begin{equation}
m_{kl}\gg 2r^{-1} \gg
m_{i_p i_q}\>, \quad \forall p,q\in\{1\ldots n\}\quad
{\rm with}\quad (p,q)\not=(1,n)\>,
\lab{SepareChain}
\end{equation}
where the numbers $m_{ij}$ are defined by eq.~\rf{Mdef}, but now as a
function of the (quantum) $S$-matrix parameters. Conversely, those
pseudoenergies do depend on each other for
\begin{equation}
2r^{-1}\gg m_{i_p i_q}\>, \quad \forall p,q\in\{1\ldots n\}\>.
\end{equation}
This way, and provided that the scale $m_{kl}$ is well-separated
from the other scales associated with this chain of nodes,
the value of $c(r)$ will change abruptly at $2r^{-1}\simeq m_{kl}$. These
changes correspond to the decoupling of all the particles (stable and
unstable) associated with the roots
$\bev$ such that both $\bev\cdot\lav_k\not=0$ and $\bev\cdot\lav_l\not=0$.
When `$k$' and `$l$' are themselves neighbouring nodes on the Dynkin
diagram of $g$, the corresponding changes of behaviour of $c(r)$
were already noticed in~\cite{HSGTBA} (see also~\cite{PARIS,FFRG}).

Therefore, the study of the TBA equations leads to the conclusion that
the form of the scaling function
$c(r)$ will be characterised by the value of the mass scales defined by
eqs.~\rf{LambdaW} and~\rf{MassRoot}, which we repeat here:
\bea
&&m^2_{\vec\beta}=(\vec\lambda_+\cdot
\vec\beta)\,(\vec\lambda_-\cdot\vec\beta)~,\nn
&&\vec\lambda_{\pm}=\sum_1^{\rg}m_i\> e^{\pm\sigma_i}\>
\vec\lambda_i~,\quad
\vec\lambda_i\cdot \vec\alpha_j=\delta_{ij}~,
\lab{PhysScales}
\ena
where $\bev$ is any positive root of $g$, and $\{m_i, \sigma_i\}$ are
the $S$-matrix parameters.
The scaling function $c(r)$ becomes nearly constant in the regions where
$2r^{-1}$ is between two consecutive mass scales, but far away from
their actual values. Equivalently, $c(r)$ develops a plateau when the
values of two of the scales provided by~\rf{PhysScales} are well-separated
and there is no other scale between them. This way, $c(r)$ is
predicted to
exhibit a staircase pattern with a finite number of plateaux, whose
maximum number depends on the Lie group $G$ and is given by
table~\ref{MaxNumb}. Then, using techniques similar to
those introduced, for instance, in~\cite{PAT}, it is possible to
calculate the value of $c(r)$ on these plateaux, and thereby map out
the RG flows. Further details will be given in \cite{NEW}. 

Finally, we remark that
for the HSG models there is an alternative way to calculate
the plateau values of $c(r)$,
using the physical interpretation of the staircase pattern
as a consequence of the decoupling of those particles that are effectively
heavy at the relative energy scale fixed by the temperature $r^{-1}$,
together with the Lagrangian description provided by~\rf{Action}.
In addition, this procedure leads to a well-defined conjecture for the
relationship between Lagrangian and $S$-matrix parameters.
Once more, we refer the interested reader to~\cite{NEW} for details.

\FIGURE[ht]{
\epsfig{file=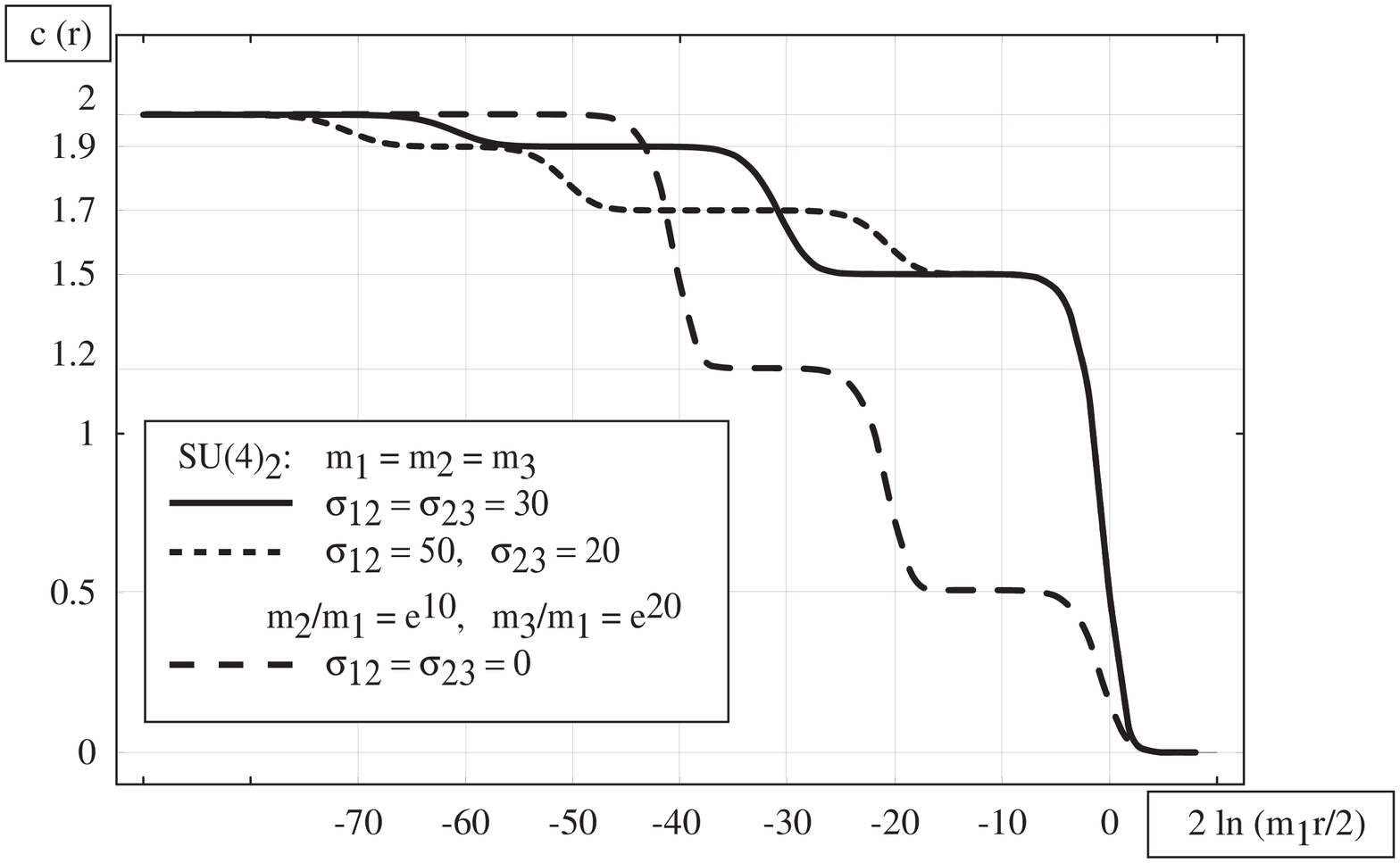,height=9truecm,width=12truecm}
\caption{The TBA scaling function for the $SU(4)_2/U(1)^3$ HSG model.}
\label{fig1}
}

The results summarised in the previous paragraphs rely on assumptions
about the behaviour of the solutions to the TBA equations, which, though
well-motivated, have not been rigorously established. Therefore, we
have subjected them to some independent checks. In particular, the TBA
equations~\rf{TBAGen} have been solved numerically in a number of
cases. Figures~\ref{fig1} and~\ref{fig2} show some numerical results
for $G=SU(4)$, $SU(5)$, and $SO(8)$, with $k=2$ and different choices for
the resonance parameters and the mass scales.
In order to illustrate the resulting patterns of flows, we have collected
some of them in the appendix. Remarkably, they can be understood as
composite flows made of elementary massive and massless flows between
conformal field theories. Results for $G=SU(3)$ and
$k=2,3,4$ were presented in~\cite{HSGTBA}. In all the cases considered,
the agreement with our predictions is excellent. In particular, they
confirm that the form of the scaling function $c(r)$ depends on the value
of the scales given by~\rf{PhysScales}, associated with all the positive
roots of $g$.

\FIGURE[ht]{
\epsfig{file=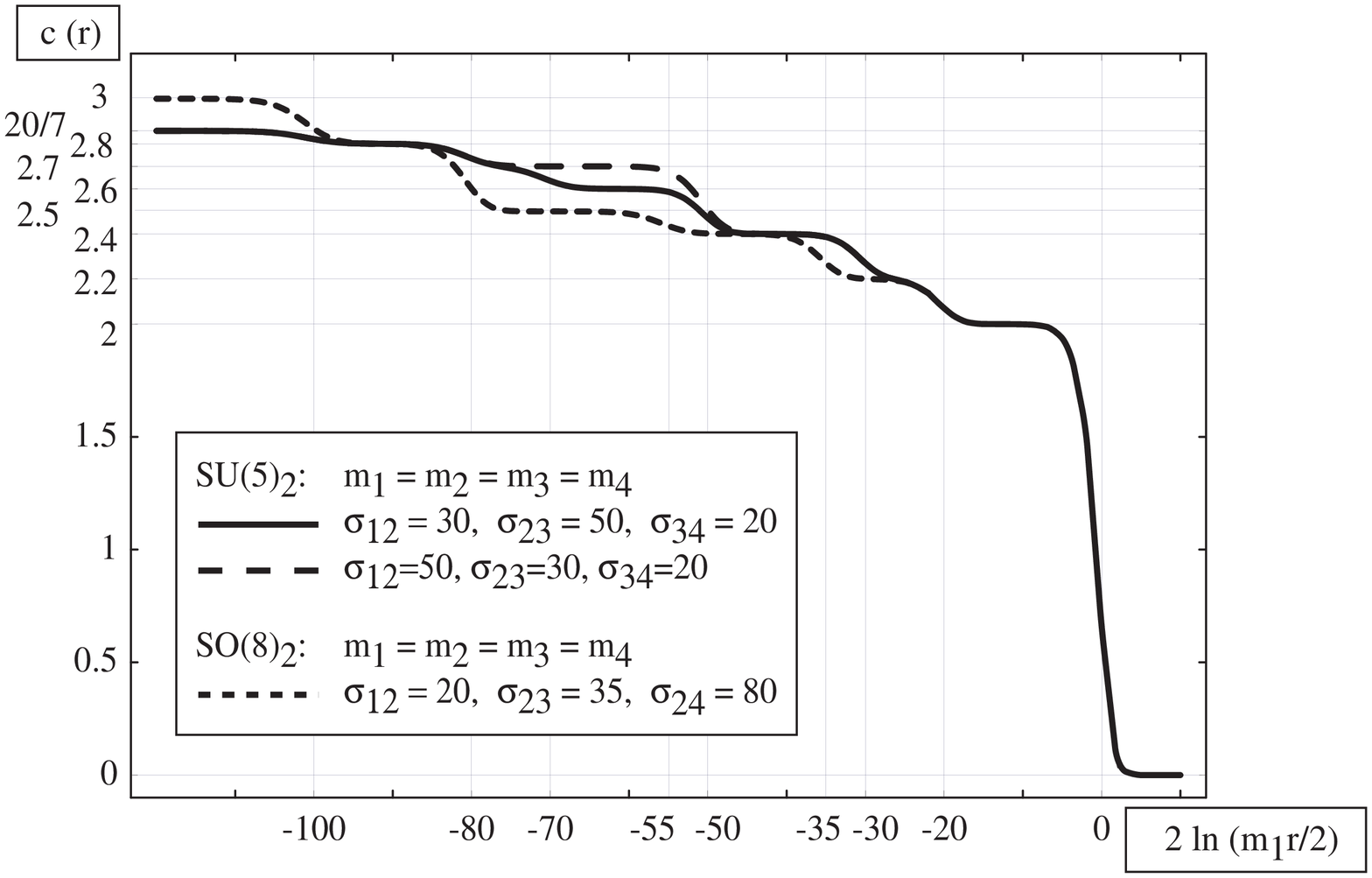,height=9truecm,width=12cm}
\caption{The TBA scaling function for the $SU(5)_2/U(1)^4$  and
$SO(8)_2/U(1)^4$ HSG models.}
\label{fig2}
}

\section{Summary and discussion}

To sum up, the main result of~\cite{NEW} is that the study of
finite-size effects using the TBA provides non-perturbative evidence for
the existence of an independent mass scale associated with each positive
root of $g$ in the
$G_k$-HSG models, although this way only well-separated scales can be
directly seen. This is in  agreement with the results of~\cite{HSGTBA}
and~\cite{FFRG} for the $SU(3)_k$--HSG models, but disagrees with the
results
of~\cite{FF2} for
$SU(N)_2$ with $N\geq4$. This work is based on the form-factor approach,
and failed to detect any of the
$(N-3)(N-2)/2$ mass scales associated with the roots of $a_{N-1}$ of
height larger than~2. 
The resolution of this discrepancy could reveal  some novel
features of the form-factor method.

Let us finish by mentioning three open questions about the HSG
models raised by our work and not addressed in this article. First of all,
and despite the fact that these models are described by diagonal
$S$-matrix theories, the
influence of the extra unstable particles gives the HSG models a
much richer structure of RG flows than had
previously been claimed, giving them a rather universal status which
unifies the simplest flows between CFTs within a common structure.
This should help towards a better understanding of the nature of the
$G_k$-HSG models, particularly at small values of $k$, where
descriptions in terms of
coupled minimal models have recently been
investigated~\cite{PASCAL}.

The second concerns the recent observation of novel 
classical breather solutions in the complex Sine-Gordon (CSG) 
model~\cite{BOWCOCK}. It turns out that these solutions can be
understood in terms of the known spectrum of the model~\cite{BOWCOCK},
but their presence was nonetheless unexpected. The CSG model describes the
$SU(2)_k$-HSG model at large $k$, and the semiclassical
spectrum of the generic $G_k$-HSG models was constructed
by embedding the CSG $SU(2)$-solitons into $G$. Therefore, the results
of~\cite{BOWCOCK} suggest that the semiclassical analysis could still hide
some more surprises, although our results confirm to a large extent the
semiclassical picture for the HSG models derived in~\cite{SEMI}.

Finally, a feature of the HSG models which has played
very little direct role in our analysis so far is the fact that they
generally break parity. It would be very interesting to pick up signs
of this in the finite-size behaviour of the theories. Probably the
best place to start looking would be in the spectrum of
excited states. The TBA technique can also be applied to excited
states \cite{BLZ,DT}, and we hope to have more to say on this issue
in the future.

%\vskip1truecm
\acknowledgments
One of us (JLM) would like to thank the organisers of the workshop for
their kind hospitality, and congratulate the IFT people on their
50$^{\rm th}$ anniversary. PED thanks the Universidad de Santiago de
Compostela for hospitality and financial support while this work was in
progress. We are grateful to Pascal Baseilhac, Peter Bowcock, Aldo Delfino and
Roberto Tateo for valuable discussions. The work of JLM is supported partially
by  Xunta de Galicia (PGDIT00-PXI-20609), and MCyT and FEDER (BFM2002-03881 and
FPA2002-01161). We also thank the EC Commission for financial support via the
FP5 Grant HPRN-CT-2002-00325.

\vskip1truecm
\appendix
\section{Staircase patterns as RG flows.}

In this appendix, we describe explicitly the flow of effective field
theories corresponding to some of the
cases whose numerical solutions were presented in figures~\ref{fig1}
and~\ref{fig2}.

\subsection{{\mathversion{bold}$SU(4)_k$}--HSG models.}

~\indent
We consider first a case with vanishing resonance parameters, and mass
scales chosen such that $m_1\ll m_2\ll m_3$.\foot{In the following, we
will use the (standard) conventions of~\cite{KAC} for numbering the nodes
of the Dynkin diagram of $g$.} Then, the scaling function
exhibits~3 plateaux corresponding to the regions
$2r^{-1}\gg m_3$ (the deep UV limit), $m_2\ll 2r^{-1}\ll m_3$, and
$m_1\ll 2r^{-1}\ll m_2$, before it reaches the massive region for
$2r^{-1}\ll m_1$ where
$c(r)$ vanishes. Within each region, $c(r)$ equals the central charge of
the following parafermionic coset CFTs:
\begin{equation}
(UV)\>\>\>\>\frac{SU(4)_k}{ U(1)^{3}} \>
\buildrel m_3\over{\hbox to 30pt{\rightarrowfill}}\>
\frac{SU(3)_k}{U(1)^{2}} \>
\buildrel m_2\over{\hbox to 30pt{\rightarrowfill}}\>
\frac{SU(2)_k}{U(1)}\>
\buildrel m_1\over{\hbox to 30pt{\rightarrowfill}}\>
{\rm Massive}\>\>\>\> (IR)\>.
\lab{Mass1}
\end{equation}

Next, let us consider a second case with non-vanishing resonance
parameters chosen such that
\begin{equation}
\sigma_{12}\gg \sigma_{23}\gg0 \quad {\rm and}\quad m_1=m_2=m_3\>,
\end{equation}
which corresponds to
\bea
&&m_{\alv_1+\alv_2+\alv_3} \simeq m_{13}\nn
&&m_{\alv_1+\alv_2}\simeq m_{12}\>, \qquad m_{\alv_2+\alv_3} \simeq
m_{23}\>,\nn 
&&m_{\alv_1}=m_{\alv_3}=m_{\alv_3}=m_1
\ena
where
\begin{equation}
m_{13} \gg m_{12} \gg m_{23} \gg m_{1}\>.
\end{equation}
The resulting flow is
\bea
\left(\frac{SU(4)_k}{U(1)^{3}}\right)^{\>\><{2}>}
\hskip-0.3cm&& \buildrel m_{13}\over{\hbox to
30pt{\rightarrowfill}}\>
\left(\frac{SU(3)_k}{SU(2)_k\otimes
U(1)}\otimes \frac{SU(3)_k}{U(1)^2}\right)^{\>\>
<\frac{19}{10}>}\allowdisplaybreaks\nn
\noalign{\vskip0.3truecm}
&&\buildrel m_{12}\over{\hbox to
30pt{\rightarrowfill}}\>
\left(\frac{SU(3)_k}{U(1)^2}\otimes
\frac{SU(2)_k}{U(1)}\right)^{\>\> <\frac{17}{10}>}\allowdisplaybreaks\nn
\noalign{\vskip0.3truecm}
&&\buildrel m_{23}\over{\hbox to
30pt{\rightarrowfill}}\> \left(\left[
\frac{SU(2)_k}{U(1)}\right]^{\otimes3}\right)^{\>\> <\frac{3}{2}>}
\buildrel m_{1}\over{\hbox to
30pt{\rightarrowfill}}\> {\rm Massive}^{\>\> <{0}>}.
\lab{SU4flow}
\ena
where the superscript $<\;>$ provides the central charge of the
corresponding coset CFT for level $k=2$, to simplify the comparison with
the numerical results presented in figures~\ref{fig1} and~\ref{fig2}.

\subsection{{\mathversion{bold}$SU(5)_k$}--HSG models.}

~\indent
First, consider the following choice of parameters
\begin{equation}
\sigma_{23}\gg \sigma_{12}\gg \sigma_{34}\gg0\quad {\rm and} \quad
m_1=m_2=m_3=m_4\>,
\end{equation}
which leads to
\bea
&&m_{\alv_1+\alv_2+\alv_3+\alv_4} \simeq m_{14}\nn
&&m_{\alv_1+\alv_2+\alv_3}\simeq m_{13}\>, \qquad m_{\alv_2+\alv_3+\alv_4}
\simeq m_{24}\>,\nn
&&m_{\alv_1+\alv_2}\simeq m_{12}\>, \qquad
m_{\alv_2+\alv_3} \simeq
m_{23}\>,\qquad 
m_{\alv_3+\alv_4} \simeq m_{34}\nn
&&m_{\alv_1}=m_{\alv_3}=m_{\alv_3}=m_{\alv_4}=m_1
\ena
where
\begin{equation}
m_{14} \gg m_{13} \gg m_{24} \gg m_{23} \gg m_{12} \gg m_{34} \gg m_{1}\;.
\end{equation}
The resulting flow is
\bea
\left(\frac{SU(5)_k}{U(1)^{4}}\right)^{\>\><\frac{20}{7}>}
&&\buildrel m_{14}\over{\hbox to
30pt{\rightarrowfill}}\> \left(\frac{SU(4)_k}{SU(3)_k\otimes U(1)}\otimes
\frac{SU(4)_k}{U(1)^{3}}\right)^{\>\> <\frac{14}{5}>}\allowdisplaybreaks\nn
\noalign{\vskip0.3truecm}
&&\buildrel m_{13}\over{\hbox to
30pt{\rightarrowfill}}\> \left( \frac{SU(4)_k}{SU(2)_k\otimes U(1)^2}
\otimes \frac{SU(3)_k}{U(1)^{2}}\right)^{\>\>
<\frac{27}{10}>}\allowdisplaybreaks\nn
\noalign{\vskip0.3truecm}
&&\buildrel m_{24}\over{\hbox to
30pt{\rightarrowfill}}\> \left(\left[
\frac{SU(3)_k}{SU(2)_k\otimes U(1)}\right]^{\otimes2}\otimes
\frac{SU(3)_k}{U(1)^{2}}\right)^{\>\> <\frac{13}{5}>}\allowdisplaybreaks\nn
\noalign{\vskip0.3truecm}
&&\buildrel m_{23}\over{\hbox to
30pt{\rightarrowfill}}\> \left(\left[\frac{SU(3)_k}{U(1)^{2}}
\right]^{\otimes2}\right)^{\>\> <\frac{12}{5}>}\allowdisplaybreaks\nn
\noalign{\vskip0.3truecm}
&&\buildrel m_{12}\over{\hbox to
30pt{\rightarrowfill}}\> \left(\frac{SU(3)_k}{U(1)^{2}}
\otimes \left[\frac{SU(2)_k}{U(1)}\right]^{\otimes
2}\right)^{\>\> <\frac{11}{5}>}\allowdisplaybreaks\nn
\noalign{\vskip0.3truecm}
&&\buildrel m_{34}\over{\hbox to
30pt{\rightarrowfill}}\> \left(\left[\frac{SU(2)_k}{U(1)}
\right]^{\otimes 4}\right)^{\>\> <{2}>}
\buildrel m_{1}\over{\hbox to
30pt{\rightarrowfill}}\> {\rm
Massive}^{\>\> <{0}>}\>.
\ena
Notice that the flow exhibits~7 plateaux, while, according
to table~\ref{MaxNumb}, the maximum number for the $SU(5)_k$-HSG models
is 10. The difference is due to the fact that we have chosen the 4 stable
mass scales to be equal, which means that~3 of them are not separated.

Second, consider another choice of parameters such that
\bea
&&\sigma_{12}\gg \sigma_{23}\gg \sigma_{34}\gg0\>, \qquad
\sigma_{23}+\sigma_{34}=\sigma_{12}\>,\nn
&& {\rm and}\quad m_1=m_2=m_3=m_4\>,
\ena
which corresponds to
\begin{equation}
m_{14} \gg m_{13} \gg m_{24}= m_{12} \gg m_{23} \gg m_{34}
\gg m_{1}\;.
\end{equation}

The resulting flow is
\bea
\left(\frac{SU(5)_k}{U(1)^{4}}\right)^{\>\><\frac{20}{7}>}
&&\buildrel m_{14}\over{\hbox to
30pt{\rightarrowfill}}\> \left(
\frac{SU(4)_k}{SU(3)_k\otimes U(1)}\otimes
\frac{SU(4)_k}{U(1)^{3}}\right)^{\>\>
<\frac{14}{5}>}\allowdisplaybreaks\nn
\noalign{\vskip0.3truecm}
&&\buildrel m_{13}\over{\hbox to
30pt{\rightarrowfill}}\> \left(\frac{SU(4)_k}{SU(2)_k\otimes
U(1)^2}\otimes \frac{SU(3)_k}{U(1)^{2}}\right)^{\>\>
<\frac{27}{10}>}\allowdisplaybreaks\nn
\noalign{\vskip0.3truecm}
&&\buildrel m_{24}=m_{12}\over{\hbox to
60pt{\rightarrowfill}}\> \left(\left[
\frac{SU(3)_k}{U(1)^{2}}
\right]^{\otimes2}\right)^{\>\> <\frac{12}{5}>}\allowdisplaybreaks\nn
\noalign{\vskip0.3truecm}
&&\buildrel m_{23}\over{\hbox to
30pt{\rightarrowfill}}\> \left(\frac{SU(3)_k}{U(1)^{2}}\otimes
\left[\frac{SU(2)_k}{U(1)}\right]^{\otimes 2}\right)^{\>\>
<\frac{11}{5}>}\allowdisplaybreaks\nn
\noalign{\vskip0.3truecm}
&&\buildrel m_{34}\over{\hbox to
30pt{\rightarrowfill}}\> \left(\left[\frac{SU(2)_k}{U(1)}
\right]^{\otimes 4}\right)^{\>\> <{2}>}
\buildrel m_{1}\over{\hbox to
30pt{\rightarrowfill}}\> {\rm Massive}^{\>\> <{0}>}\>.
\lab{SU5FlowB}
\ena
This case exhibits one plateau less that the previous one,
which is a consequence of the fact that the scales $m_{24}$ and $m_{12}$
are not separated for this choice of parameters.

\subsection{{\mathversion{bold}$SO(8)_k$}--HSG models.}

~\indent
We shall only consider the choice of parameters
\begin{equation}
\sigma_{12}=20\>, \quad \sigma_{23}=35\>,\quad \sigma_{24}=80\>, \quad{\rm
and}\quad m_1=m_2=m_3=m_4\>,
\end{equation}
which corresponds to
\bea
&&m_{\alv_1+2\alv_2+\alv_3+\alv_4}\>,\;
m_{\alv_1+\alv_2+\alv_3+\alv_4}\>,\; m_{\alv_1+\alv_2+\alv_4}\simeq
m_{14}\nn  &&m_{\alv_2+\alv_4}\>,\; m_{\alv_2+\alv_3+\alv_4}\simeq
m_{24}\>, \nn 
&&m_{\alv_1+\alv_2+\alv_3}\simeq m_{13}\>, \nn
&&m_{\alv_2+\alv_3}\simeq m_{23}\>, \quad m_{\alv_1+\alv_2}\simeq
m_{12}\>, \nn 
&&m_{\alv_1}=m_{\alv_3}=m_{\alv_3}=m_1
\ena
where
\begin{equation}
m_{14} \gg m_{24} \gg m_{13} \gg m_{23} \gg m_{12} \gg m_{1}\;.
\end{equation}
The resulting flow is
\bea
\left(\frac{SO(8)_k}{U(1)^{4}}\right)^{\>\><{3}>}
&&\hskip-0.3cm\buildrel m_{14}\over{\hbox to
30pt{\rightarrowfill}}\>
\left(\frac{SU(4)_k}{SU(3)_k\otimes
U(1)}\otimes \frac{SU(4)_k}{U(1)^3}\right)^{\>\>
<\frac{14}{5}>}\allowdisplaybreaks\nn
\noalign{\vskip0.3truecm}
&&\hskip-0.3cm\buildrel m_{24}\over{\hbox to
30pt{\rightarrowfill}}\>  \equiv
\left(\frac{SU(4)_k}{U(1)^3}\otimes
\frac{SU(2)_k}{U(1)}\right)^{\>\> <\frac{5}{2}>}\allowdisplaybreaks\nn
\noalign{\vskip0.3truecm}
&&\hskip-0.3cm\buildrel m_{13}\over{\hbox to
30pt{\rightarrowfill}}\> \left(\frac{SU(3)_k}{SU(2)_k\otimes U(1)}
\otimes \frac{SU(3)_k}{U(1)^{2}}
\otimes \frac{SU(2)_k}{U(1)}
\right)^{\>\> <\frac{12}{5}>}\allowdisplaybreaks\nn
\noalign{\vskip0.3truecm}
&&\hskip-0.3cm\buildrel m_{23}\over{\hbox to
30pt{\rightarrowfill}}\> \left(\frac{SU(3)_k}{U(1)^{2}}\otimes
\left[\frac{SU(2)_k}{U(1)}\right]^{\otimes2}\right)^{\>\>
<\frac{11}{5}>}\allowdisplaybreaks\nn
\noalign{\vskip0.3truecm}
&&\hskip-0.3truecm\buildrel m_{12}\over{\hbox to
30pt{\rightarrowfill}}\> \left(\left[\frac{SU(2)_k}{U(1)}
\right]^{\otimes4}\right)^{\>\> <{2}>}
\buildrel m_{1}\over{\hbox to
30pt{\rightarrowfill}}\>
{\rm Massive}^{\>\> <{0}>}.
\lab{SO8flow}
\ena
In this case the flow exhibits~6 plateaux, which agrees with the
maximal number quoted in table~\ref{MaxNumb}, which is~9, taking into
account that we have chosen the 4 stable mass scales to be equal.

\end{document}